\documentclass[12pt,preprint]{aastex}
\usepackage{emulateapj5}

\begin {document}

\title{Discovery of a New Dusty B[e] Star in the Small Magellanic Cloud}

\author{John P. Wisniewski\altaffilmark{1,2,3}, 
Karen S. Bjorkman\altaffilmark{3,4}, Jon E. Bjorkman\altaffilmark{4}, 
Mark Clampin\altaffilmark{2}}

\altaffiltext{1}{NPP Fellow}
\altaffiltext{2}{NASA GSFC Exoplanets and Stellar Astrophysics Lab Code 667, 
Greenbelt, MD 20771, John.P.Wisniewski@nasa.gov, Mark.Clampin@nasa.gov}
\altaffiltext{3}{Visiting Astronomer, Cerro Tololo Inter-American Observatory} 
\altaffiltext{4}{Ritter Observatory, MS \#113, Department of Physics
and Astronomy, University of Toledo, Toledo, OH 43606, Karen.Bjorkman@utoledo.edu, 
Jon.Bjorkman@utoledo.edu}

\begin{abstract}

We present new optical spectroscopic and archival Spitzer IRAC photometric observations 
of a B-type star in the SMC cluster NGC 346, NGC 346:KWBBe 200.  We detect 
numerous Fe II, [O I], and [Fe II] lines, as well as strong P-Cygni profile H I 
emission lines in its optical spectrum.  The star's near-IR color and optical to IR SED 
clearly indicate the presence of an infrared excess, consistent with the presence 
of gas and warm, T $\sim$800 K, circumstellar dust.  Based on a crude 
estimate of the star's 
luminosity and the observed spectroscopic line profile morphologies, we find 
that the star is likely to be a B-type supergiant.  We suggest that 
NGC 346:KWBBe 200 is a newly discovered B[e] supergiant star, and represents 
the fifth such object to be identified in the SMC.

\end{abstract}


\section{Introduction} \label{introduction}

Massive OB stars play an important role in astrophysics; their winds and ejecta 
shape the morphology and ionization structure of their local environment, and provide 
a source of heavy elements which serve as the building blocks of planetary systems.  
B[e] stars comprise a heterogeneous group of massive stars which all exhibit similar 
observational properties, yet range in evolutionary age from the pre- to post-main 
sequence \citep{lam98}.  B[e] supergiants (B[e]SGs) represent one 
sub-class of the B[e] phenomenon and are notable in that at least some share similar 
photometric, spectroscopic, and rotational velocity properties as Luminous Blue 
Variables (LBVs) \citep{sta83,zic96,van02,zic06}, which suggests B[e]SGs might be 
precursors to LBVs.

B-type stars which exhibit strong H I Balmer emission, forbidden and low excitation 
emission lines, commonly Fe II, [Fe II], and [O I], and evidence of warm (T$\sim$ 1000 K) 
dust are assigned the classification of B[e] \citep{all76,lam98,zic06}.  
B[e]SGs are known to exhibit a hybrid spectrum characterized by broad UV resonance 
absorption lines and a wealth of narrow emission lines \citep{zic85,zic86}.  \citet{zic85} 
suggested that a non-spherical wind model consisting of a cool, equatorial component, 
and a hot, polar component best explained these observations.  Subsequent polarimetric 
observations seem to confirm the presence of an axisymmetric circumstellar 
environment in B[e]SGs \citep{mag92,sch93,mel01}, the base of which may sometimes be 
clumpy \citep{mag06}.  While polarimetric observations indicate the presence of an asymmetric, 
clumpy wind in some LBVs and LBV-like objects \citep{tay91,nor01,dav05,wis07}, it is 
not yet clear whether they exhibit a clumpy, axi-symmetric geometry similar to 
that observed for B[e] stars (see e.g. \citealt{dav05}), which might be expected if B[e]SGs are 
truly precursors of LBVs.   

To date, four B[e] stars have been identified in the Small Magellanic Cloud (SMC), 
while eleven have been identified in the Large Magellanic Cloud (LMC).  Owing to the known distance of the SMC and LMC and the observed luminosities of the fifteen SMC and LMC B[e] sources, it is well accepted that all are B[e]SGs \citep{zic06}.  In this paper, we report the serendipitous discovery of the fifth known B[e] star in the SMC.
The star was initially classified by \citet{mey93} as a compact H II source ([MA93] 1116) likely characterized by broad H$\alpha$ emission.  A later optical photometric survey by 
\citet{kel99} classified the same target (NGC346:KWBBe 200) as a classical Be star residing in the SMC cluster NGC 346, which is known to be a rich source of current and recent star formation \citep{not06,sab07}.  Our observations of the source were 
obtained in the context of assessing the formation and evolution of 
classical Be stars in NGC 346 and other clusters in the SMC and LMC \citep{wis06,wis07}.

\section{Observations} \label{obs}

Our spectroscopic observations were obtained on 15 November 2004 using the Cerro Tololo 
Inter-American Observatory (CTIO)\footnote{The Cerro Tololo Inter-American Observatory is operated by 
the Association of Universities for Research in Astronomy, under contract with the National Science 
Foundation.} 4m telescope Hydra 138 fiber multi-object spectrograph. 
  We used the 527 lines mm$^{-1}$ KPGL3 grating and a 200$\mu$m slit plate, yielding a dispersion of 1.75 \AA\ at 5500 \AA\ (R $\sim$3000).  We obtained 
two 3600 second exposures with NGC 346:KWBBe 200 on a fiber; for each 
exposure, we 
also assigned fourteen fibers to locations void of stellar sources to serve as probes 
of background sky emission.  Our NGC 346 exposures were bracketed by 
observations of a 
penray lamp exposure to wavelength calibrate our data.  Standard bias frames and dome 
flats were obtained at the beginning of each night of our run; sky flats were also obtained to calibrate the fiber-to-fiber throughput of Hydra.

Following bias subtraction using standard IRAF\footnote{IRAF is distributed by the National Optical Astronomy 
Observatories, which are operated by the Association of Universities for 
Research in Astronomy, Inc., under contract with the National Science 
Foundation.} techniques, we used Pieter van Dokkum's 
\textit{L.A.Cosmic} routine to identify and remove cosmic ray artifacts from our images.  Remaining 
standard reduction steps, including flat fielding, aperture extraction, and wavelength 
calibration were done within the \textit{Hydra} IRAF package.  To subtract the strong 
nebular lines present in our data, we evaluated the use of various combinations of our fourteen 
sky fibers in an iterative manner.  Following our best subtraction, 
we only see evidence of residual nebular 
contamination from [O III] lines at 4959 and 5007 \AA, at an intensity 
of 2\% of the total sky emission in these transitions.  We do not observe other narrow 
(GFWHM $\sim$2.9 \AA) emission profiles at other wavelengths, which would be 
indicative of a nebular origin; hence, we estimate that any residual nebular components 
still present in our data should contribute $<$2\% to the equivalent widths quoted in this paper.

We also present archival \textit{Spitzer Space Telescope} Infrared Array 
Camera (IRAC; \citealt{faz04}) 3.6, 4.5, 5.8, and 8 $\mu$m observations of 
NGC 346:KWBBe 200, obtained as part of program 63 (Houck PI).  The data 
were reduced via the Spitzer pipeline S14.0 software.  Photometry was extracted 
from post-pipeline mosaic images using standard IRAF \textit{DAOPHOT} 
aperture photometry techniques; uncertainties in the absolute flux calibration 
are less than 5\% \citep{ira06}.

\section{Results} \label{results}
\subsection{Optical Spectroscopy} \label{spec}

\begin{figure}
\includegraphics[scale=0.5]{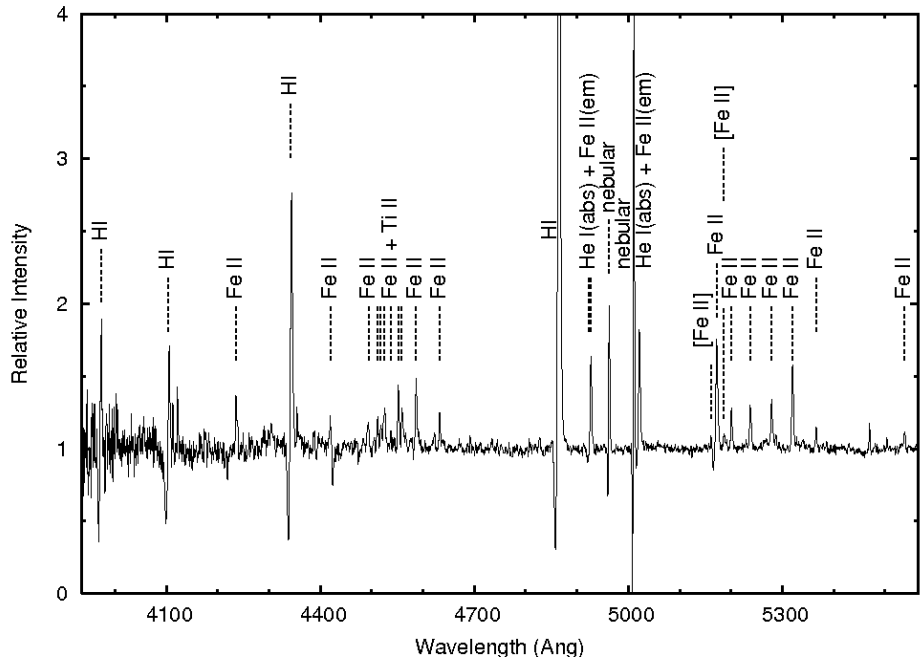} 
\caption[f1.eps]{The blue portion of NGC 346:KWBBe 200's optical spectrum is 
dominated by Fe II, [Fe II], and H I emission lines.  These transitions are typically 
observed in B[e] stars \citep{all76,lam98,zic06}. \label{blue}}
\end{figure}

\begin{figure}
\includegraphics[scale=0.5]{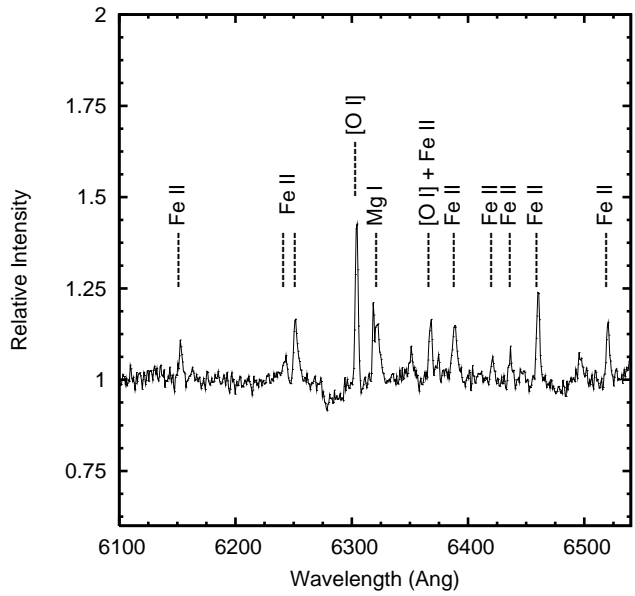} 
\caption[f2.eps]{The red portion of NGC 346:KWBBe 200's optical spectrum is 
dominated by Fe II and [O I] emission lines, which along with the transitions observed 
in the blue portion of the spectrum (Figure \ref{blue}), are typically 
observed in B[e] stars \citep{all76,lam98,zic06}. \label{red}}
\end{figure}

The continuum normalized optical spectrum of NGC 346:KWBBe 200 
from 3935-5365 \AA\ and 6000-6540 \AA\ is presented in Figures \ref{blue} and  
\ref{red} respectively.  Continuum signal to noise levels range 
from $\sim$20 at 4200 \AA\, to $\sim$70 at 
5200 and 6200 \AA.  NGC 346:KWBBe 200 is clearly characterized by a wealth of 
emission lines; we used line identifications reported by 
\citet{zic89,gum95,dew97,lam98} and \citet{mir05} to identify 
the strongest lines detected 
at the resolution of our data, which are labeled in the figures and tabulated in 
Table \ref{lines}.  The dominant species seen in emission is Fe II, although lines 
of [Fe II], [O I], Mg I, and Ti II also are present.  

\begin{figure}
\includegraphics[scale=0.3]{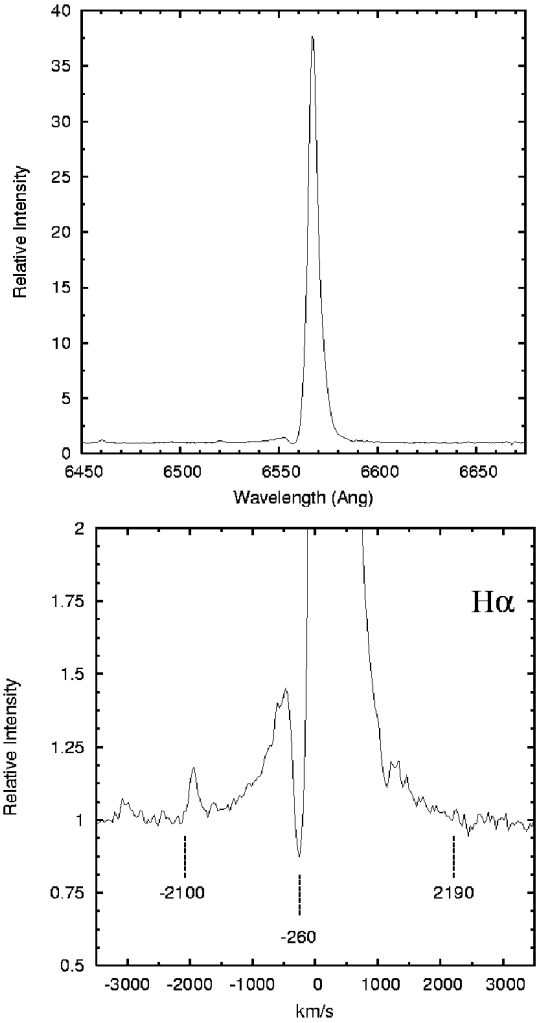}
\caption[f3.eps]{The net emission strength ($\sim$ -267 \AA), line-profile morphology (P-Cygni), 
and broad electron scattering wings ($\sim$ -2100 km s$^{-1}$ and $\sim$ +2190 km s$^{-1}$) 
of the H$\alpha$ line in NGC 346:KWBBe 200 is similar to that observed in other 
B[e]SGs \citep{zic89}.
\label{ha}}
\end{figure}

NGC 346:KWBBe 200 is also characterized by strong H I emission, from 
H$\epsilon$ (Figure \ref{blue}) through H$\alpha$ (Figure \ref{ha}), 
with all lines clearly exhibiting 
P-Cygni profiles indicative of a strong outflowing wind.  The H$\alpha$ line is particularly 
strong and broad; we measured our second exposure to have a net equivalent width 
of $\sim$ -267 \AA\ and wings extending to -2100 km s$^{-1}$ and 
+2190 km s$^{-1}$.  Similarly, we measured H$\alpha$ in our first exposure  
to have a net equivalent width of -266 \AA\ and wings extending to 
-2100 km s$^{-1}$ and +2100 km s$^{-1}$.  The dominant source of 
uncertainty in these measurements is not photon statistics, but rather 
 given the 
presence of extended electron scattering wings and the curvature of 
the spectral response of the grating+detector near H$\alpha$, our choice of 
continuum placement.  Spectra 
of emission- and non-emission line stars located in other Hydra fibers were used to 
calibrate the typical spectral response behavior near H$\alpha$, providing a constraint 
on the appropriate order of fitting function to use to fit the continuum region near 
H$\alpha$ for NGC 346:KWBBe 200.

\subsection{Optical and IR Photometry} \label{phot}

\begin{figure}
\includegraphics[scale=0.45]{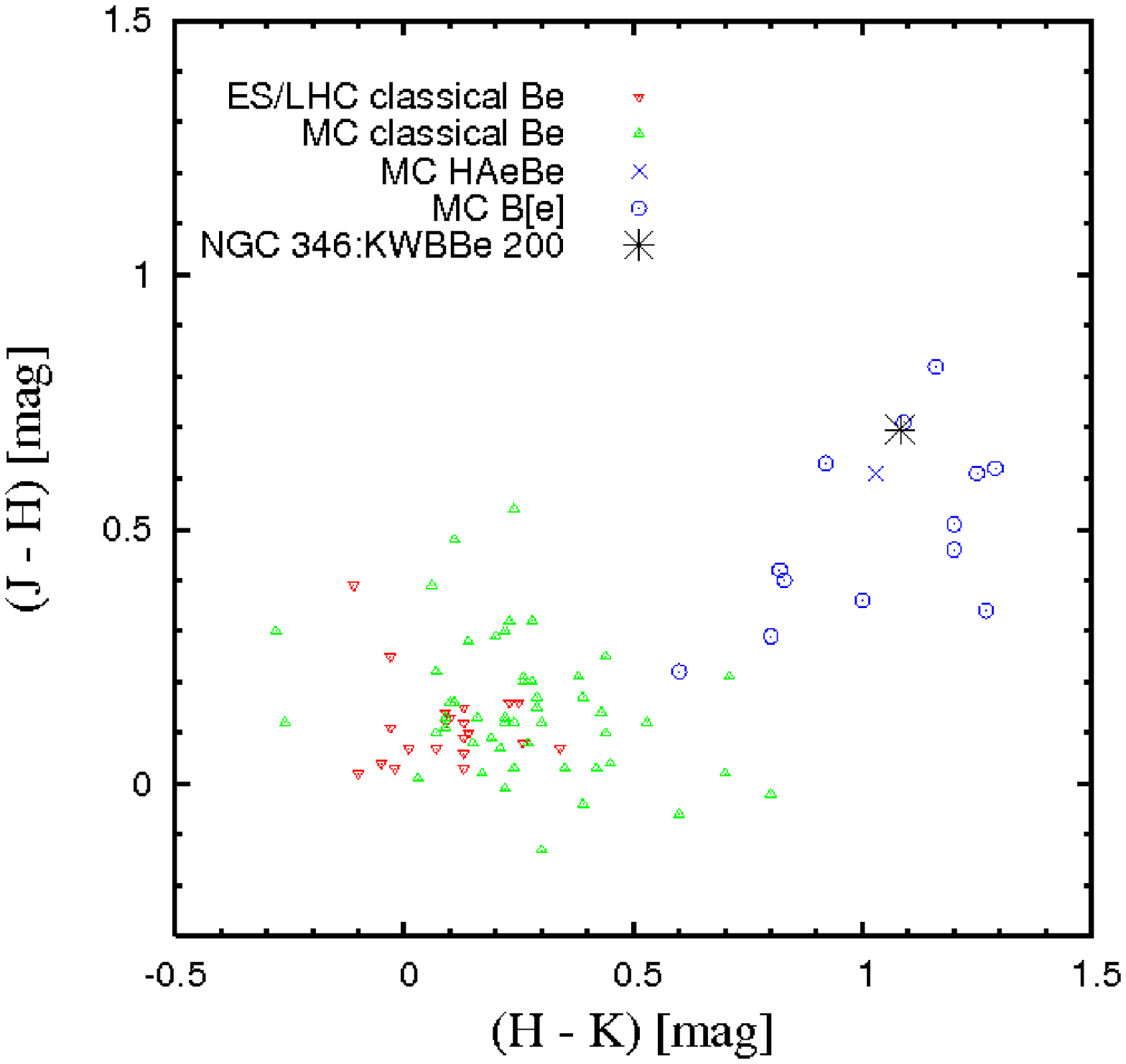} 
\caption[f4.eps]{The location of NGC 346:KWBBe 200 (black star), along with 
likely Magellanic Cloud classical Be stars from \citet{dew05} (red triangles) and 
\citet{wis07} (green triangles), a potential Herbig Ae/Be star (blue cross) in the LMC 
\citep{dew05}, and known Magellanic Cloud B[e]SGs (blue circles) from \citet{mcg88}, 
\citet{gum95}, and the 2MASS catalog are shown on a near-IR 2-color diagram. 
NGC 346:KWBBe 200's near-IR color clearly coincides with that 
of other dusty Magellanic Cloud (B[e] and Herbig Ae/Be) stars and is inconsistent 
with the observed near-IR colors of Magellanic Cloud classical Be stars. \label{2cd}}
\end{figure}

To further explore the nature 
of NGC 346:KWBBe 200, we plot the star on a near-IR 
2-color diagram (black star; Figure \ref{2cd}), along with likely 
Magellanic Cloud classical Be stars from \citet{dew05} (red triangles) and 
\citet{wis07} (green triangles), a 
potential LMC Herbig Ae/Be star (blue cross) from \citet{dew05}, and known 
SMC and LMC B[e]SGs (blue circles) from \citet{mcg88}, \citet{gum95}, and the 2MASS 
catalog.  
NGC 346:KWBBe 200's near-IR color is clearly consistent with that observed for 
other SMC/LMC B-type stars characterized by dusty circumstellar envelopes, and 
is inconsistent with the typical color of classical Be stars, which are characterized 
by gaseous circumstellar disks \citep{por03}.  

\begin{figure}
\includegraphics[scale=0.5]{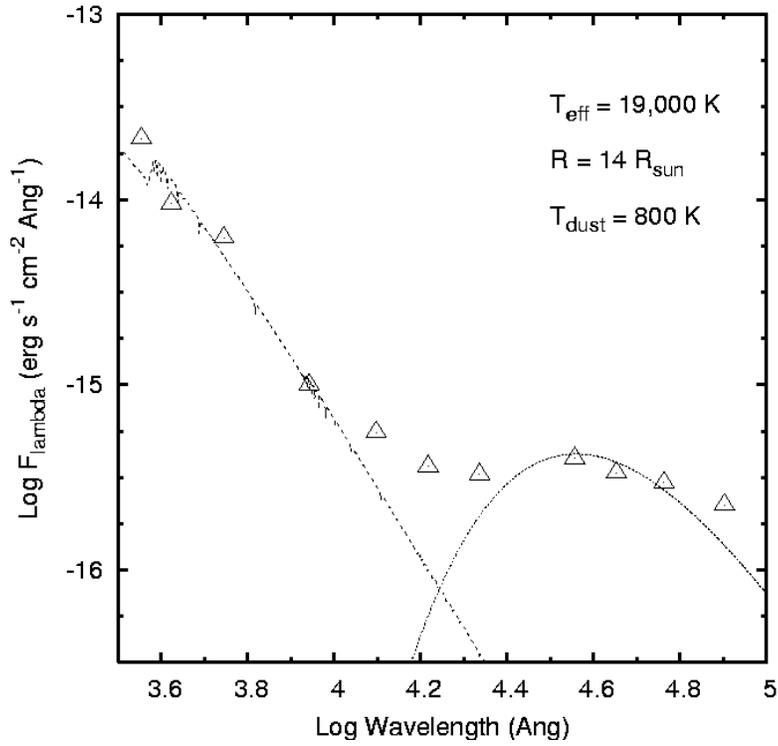} 
\caption[f5.eps]{The SED of NGC 346:KWBBe 200 following dereddening of its 
optical and near-IR (J,H,K) photometry, assuming E(B-V) = 0.35.  An excess of infrared 
flux is clearly present, indicating the presence of gas and 
warm, T$_{dust}$ $\sim$ 800K, dust.  
The optical photometry are consistent with a Kurucz model atmosphere having 
T$_{eff}$ = 19,000K and R$_{star}$ = 14 R$_{sun}$.  The extrapolated stellar 
luminosity, log (L/L$_{sun}$) $\sim$4.4, and observed line profile morphologies 
suggest that NGC 346:KWBBe 200 is a B[e]SG. \label{sed}}
\end{figure}

We next construct NGC 346:KWBBe 200's spectral energy distribution (SED), using all available photometry.  Its observed (B-V) and (V-I) colors, 
0.351 and 0.366 \citep{zar02}, are substantially redder than the expected intrinsic 
colors of main sequence (-0.30 $<$ (B-V)$_{o}$ $<$ -0.07, \citealt{sch82}; -0.44 $<$ 
(V-I)$_{o}$ $<$ -0.11, \citealt{duc01}) or supergiant-type (-0.23 $<$ (B-V)$_{o}$ $<$ 
0.0, \citealt{sch82}; -0.37 $<$ (V-I)$_{o}$ $<$ 0.0, \citealt{duc01}) B stars.  As such, 
we have assumed that NGC 346:KWBBe 200 is characterized by a \textit{minimum} 
E(B-V) reddening of 0.35, and de-reddened its U- through K-band photometry 
using R$_{V}$ = 2.74 \citep{gor03} and the standard extinction curves of 
\citet{car89}.  The resultant SED incorporating U-band through Spitzer IRAC 8 $\mu$m photometry is shown in Figure \ref{sed}.  An excess of IR emission is evident both at 
near-IR (J-band) wavelengths, likely originating from free-free and bound-free emission
from hydrogen in 
a circumstellar envelope (or disk), and at IRAC-band IR wavelengths, indicating the presence of warm dust in a circumstellar envelope (or disk) 
(see e.g. \citealt{zic85,zic89}).  To crudely characterize this dust component, 
we have overlayed a Planck function corresponding to a dust temperature of 
T = 800K in Figure \ref{sed}, which we believe best represents the observed SED trend.

\section{Discussion}
\subsection{NGC 346:KWBBe 200's Status as a B[e] Star}

We have shown that NGC 346:KWBBe 200's optical spectrum is dominated by 
emission lines from Fe II, [O I], and [Fe II].  It exhibits broad P-Cygni 
profile H I emission lines, indicative of a strong outflowing wind; the strength of its   
H$\alpha$ equivalent width, $\sim$ -267 \AA\, is rivaled only by other 
known B[e]SGs \citep{zic89} and some LBVs \citep{kda05}.  The star's 
circumstellar environment is clearly characterized by the presence of both gas 
and warm dust, as diagnosed from its near-IR colors (Figure \ref{2cd}) and near- and 
IR-SED excess (Figure \ref{sed}).  These observational properties are characteristic 
of B[e] stars \citep{all76,lam98,zic06}, and inconsistent with the expected behavior of classical Be stars; thus, we suggest that NGC 346:KWBBe 200 should be re-classified 
as the fifth known B[e] star in the SMC.

\subsection{Evolutionary Status}

To constrain the evolutionary status of NGC 346:KWBBe 200, we compared 
its de-reddened photometry to standard Kurucz model atmospheres \citep{kur92} using 
log (Z/Z$_{sun}$) = -1.0 and log (g) = 3.5.  
As shown in Figure \ref{sed} and summarized in Table \ref{param}, the SED data are 
best represented by a model having T$_{eff}$ = 19,000 K, and R$_{star}$ = 14 
R$_{sun}$ (Figure \ref{sed}).  Given the significant 
uncertainty over the exact E(B-V) reddening associated with NGC 346:KWBBe 200, we 
caution that these stellar parameters should only be considered initial estimates; 
the presence of a higher E(B-V) reddening than assumed here (0.35) will inflate T$_{eff}$ 
and modify R$_{star}$.  Nonetheless, assuming a standard relationship that 
L$_{star}$ = (R$_{star}$/R$_{sun}$)$^{2}$ (T$_{eff}$/T$_{sun}$)$^{4}$), we can 
estimate the luminosity of NGC 346:KWBBe 200 to be log (L/L$_{sun}$) $\sim$4.4 
(Table \ref{param}).  These stellar parameters correspond to a crude spectral classification 
of B3[e] II.  

\begin{figure}
\includegraphics[scale=0.5]{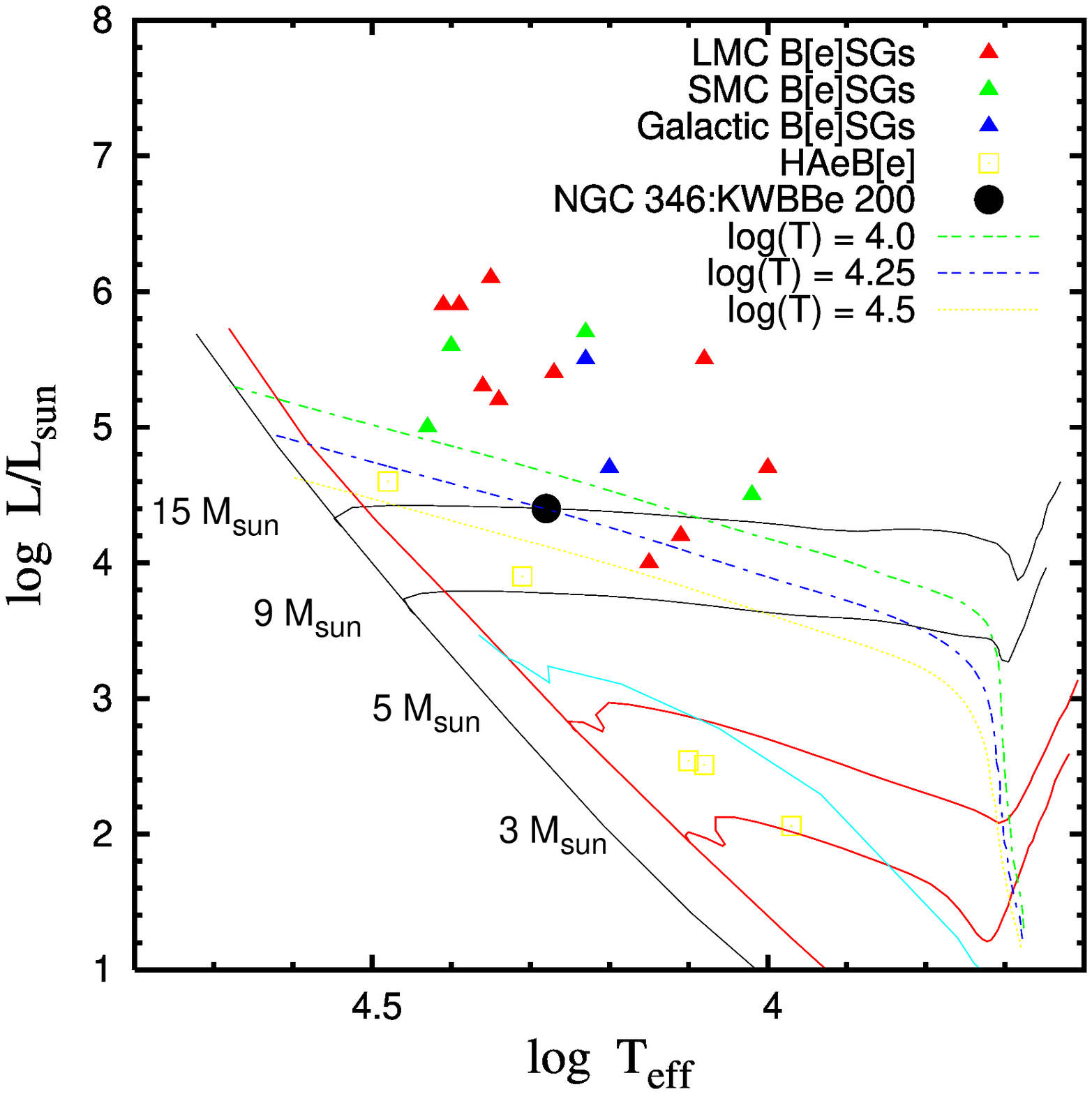} 
\caption[f6.eps]{NGC 346:KWBBe 200 (black circle) is plotted on a HR diagram, along with 
other known LMC, SMC, and Galactic B[e]SGs (red, green, and blue triangles respectively) 
and Galactic HAeBe stars (yellow squares).  Temperature and luminosity values 
were taken from \citet{lam98} and references therein for all of these ancillary sources, except 
for the luminosity of MWC 1080 \citep{hen98} and V594 Cas \citep{vin07}, whose tabulated values 
within \citet{lam98} did not match their quoted literature origin and/or could not be otherwise verified.
Also overplotted are the canonical evolutionary tracks and ZAMS for 9 and 15 M$_{sun}$ (z = 0.001) 
stars (black lines) and 3 and 5 M$_{sun}$ (z = 0.020) stars (red lines), the birthline for the z = 0.020 
models (light blue line), and isochrones for the z = 0.001 models (dashed green, blue, and 
yellow lines).  NGC 346:KWBBe 200 is well above the birthline of late-type B stars (5 M$_{sun}$), 
and appears to be most similar to the lower luminosity 
Magellanic Cloud B[e]SGs stars reported by \citet{gum95}.  Furthermore, its observed photometric 
and spectroscopic properties do not 
indicate that the star is highly embedded in a natal star formation envelope, as would be expected 
if it were truly located at the log(T) = 4.25 isochrone of the 15 M$_{sun}$ pre-MS evolutionary 
track of \citep{ber96}.  \label{hrd}}
\end{figure}

Using these crude stellar parameters, we plot NGC 346:KWBBe 200 on a Hertzsprung-Russell (HR) 
diagram (black circle), along with other known SMC, LMC, and Galactic B[e]SG (green, red, and blue 
triangles) and pre-main sequence Herbig B[e] (hereafter HAeB[e]) stars (yellow squares) in Figure \ref{hrd}.  
The canonical (non-accreting) evolutionary tracks \citep{ber96} and 
zero-age main sequence (ZAMS) for 9 and 15 M$_{sun}$ stars at a metallicity appropriate for the SMC/LMC, 
z = 0.001 (black lines), and for 3 and 5 M$_{sun}$ stars at a metallicity appropriate for the Galactic HAeBe 
stars, z = 0.020 (red lines), are also plotted in Figure \ref{hrd}, along with the birthline for the z = 0.020 
models (solid light blue line) and isochrones for the z = 0.001 models (dashed green, blue, and yellow lines).
It is clear that NGC 346:KWBBe 200 is well above the birthline of late-type B stars (5 M$_{sun}$) and 
is most similar to the lower luminosity 
Magellanic Cloud B[e]SGs stars reported by \citet{gum95}.  If it were a pre-MS star, inspection of 
Figure \ref{hrd} indicates that NGC 346:KWBBe 200 would 
be positioned at a youthful position along the the z = 0.001 
15 M$_{sun}$ evolutionary track, and intersect the log(T) = 4.25 isochrone.  Such a young, high-mass
object would likely still be deeply embedded in its natal star formation envelope, in contradiction to the 
star's observed optical V-band magnitude and our easy detection of its optical spectrum.

\citet{lam98} provided a nice discussion of the classification of B[e] stars and established numerous criteria for 
determining whether a B[e] star is a B[e]SG, HAeB[e], compact planetary nebula B[e], Symbiotic B[e], or an 
``unclassified'' B[e] star, which we briefly review here.  To be considered a B[e]SG, \citet{lam98} suggest 
stars should obey two primary criteria: A1) exhibit the B[e] phenomenon; and A2) be supergiants with 
log (L/L$_{sun}$) $\geq$ 4.0; and four secondary criteria, two of which are: B1) optical spectroscopic 
evidence of mass loss; and B2) typically exhibit small photometric variations of order 0.1$^{m}$.  
To be considered a HAeB[e], \citet{lam98} suggest stars should obey three 
primary criteria: A1) exhibit the B[e] phenomenon; A2) be associated with a star forming region; and A3) 
exhibit evidence of accretion via inverse P Cygni spectroscopic line profiles; and three secondary 
criteria: B1) exhibit a luminosity of log (L/L$_{sun}$) $\leq$ 4.5; B2) exhibit large, irregular photometric 
variations; and B3) exhibit evidence of warm and cool dust in their SEDs.  Note that some of the criteria 
for B[e]SG and HAeB[e] designations overlap; moreover, a growing body of evidence suggests that some 
B[e]SGs may also exhibit significant, large amplitude photometric variability \citep{zic96,van02}, indicating 
that this behavior is not exclusive to HAeB[e] stars.

NGC 346:KWBBe 200's strong P-Cygni profile H I emission is indicative of mass-loss; in our spectrum, 
we observe no evidence 
of inverse P-Cygni profiles which would be indicative of infall.  As shown in Figure \ref{hrd}, NGC 346:KWBBe 200's 
crudely derived luminosity, log (L/L$_{sun}$) $\sim$4.4, and location on a HR diagram is more consistent with 
the star being a low luminosity B[e]SG than a high luminosity HAeB[e] star.  The star does 
reside in a young cluster, portions of which are actively forming low-mass stars \citep{not06,sab07}, which 
seemingly agrees with the aforementioned ``A2'' criteria of HAeB[e] stars; however, other well known 
highly evolved massive stars reside in the region (e.g. the Wolf-Rayet/LBV HD 5980; aka 
NGC 346:KWBBe 13 \citealt{kel99}) hence this particular criteria is probably not a useful discriminant in 
this situation.  As such, given the available data presently available for NGC 346:KWBBe 200, we suggest the 
star is a B[e]SG, similar to the other 4 known SMC and 11 LMC B[e] identified to date \citep{gum95,lam98}.

\section{Summary and Future Work}

NGC 346:KWBBe 200 is a B-type star whose optical spectrum is dominated by Fe II, [O I], [Fe II], and strong P-Cygni H I emission lines, and exhibits clear evidence of having a 
circumstellar envelope characterized by the presence of gas and warm 
(T$_{dust}$ $\sim$800 K) dust.  Based on these observational properties, we suggest that 
NGC 346:KWBBe 200 is a B[e]SG star, representing the fifth such object identified to date in the SMC.  Our 
crude estimate of the star's luminosity, log (L/L$_{sun}$) $\sim$4.4, its location on a HR diagram, and its 
observed line profile morphologies suggests it is most likely to be a B[e] supergiant.

We recommend several observational approaches be pursued to further constrain the evolutionary 
status of NGC 346:KWBBe 200.  Mid- to far-IR photometric observations would allow one to search 
for the presence of cool dust, which is one of the defining characteristics of HAeB[e] stars \citep{lam98} 
and would not be an expected 
characteristic of B[e]SGs.  Moreover, high resolution optical and UV spectroscopic observations would 
facilitate the derivation of a more reliable estimate of NGC 346:KWBBe 200's spectral type and luminosity, 
hence providing more conclusive evidence of the post-main sequence 
evolutionary status suggested in this paper.  Optical photometric monitoring, to search for and characterize 
variability, and measurements of the star's rotational velocity would aid efforts to explore evolutionary 
links between the B[e]SG and LBV phases of massive star evolution.

\acknowledgments

We thank Ted Gull and Aki Roberge for helpful discussions about these results.  We also thank our referee, 
Jorick Vink, for providing useful feedback which improved the content and presentation of this paper.
Support for this project was provided by NASA NPP and GSRP fellowships to JPW (NNH06CC03B, NGT5-50469), a NASA LTSA grant NAG5-8054 and a Research Corporation Cottrell Scholar award to KSB, and a NSF grant (AST-0307686) to JEB.  
This work is based in part on observations made with the Spitzer Space Telescope, 
which is operated by the Jet Propulsion Laboratory, California Institute of 
Technology under a contract with NASA.  We have also made use of the SIMBAD database operated at CDS, Strasbourg, France, and the NASA ADS system.

\clearpage

\clearpage

\clearpage

\clearpage

\clearpage

\clearpage

\begin{table} 
\caption{Summary of Observed Emission and Absorption Features \label{lines}}
\scriptsize
\begin{tabular}{lccccc}
Wavelength & Line & EW (\AA) & comment \\
\tableline
\\
3970 & H $\epsilon$ & \nodata & P-Cygni \\
4101 & H $\delta$ & 0.4 & P-Cygni \\
4233 & Fe II & -1.3 & \nodata \\
4340 & H $\gamma$ & -5.2 & P-Cygni \\
4417 & Fe II & \nodata & \nodata \\
4491 & Fe II & \nodata & blend$^{1}$  \\
4508 & Fe II & \nodata & \nodata \\
4515 & Fe II & \nodata & \nodata \\
4522 & Fe II & \nodata & blend$^{2}$  \\
4534 & Ti II + Fe II & \nodata & blend$^{3}$  \\
4549 & Fe II + Ti II & \nodata & blend$^{4}$  \\
4555 & Fe II & \nodata & \nodata \\
4584 & Fe II & \nodata & \nodata \\
4629 & Fe II & \nodata & \nodata \\
4861 & H $\beta$ & -29.3 & P-Cygni$^{5}$  \\
4921 & He I & \nodata & abs \\
4924 & Fe II & -2.3 & \nodata \\
5015 & He I & \nodata & abs \\
5018 & Fe II & -3.5 & \nodata \\
5158 & [Fe II] & \nodata & \nodata \\
5169 & Fe II & \nodata & P-Cygni? \\
5184 & [Fe II] & \nodata & \nodata  \\
5198 & Fe II & -1.0 & \nodata \\
5235 & Fe II & -1.2 & \nodata \\
5276 & Fe II & \nodata & \nodata \\
5317 & Fe II & -2.4 & \nodata \\
5363 & Fe II & \nodata & \nodata \\
5535 & Fe II & \nodata & \nodata \\
6148 & Fe II & -0.4 & \nodata \\
6238 & Fe II & \nodata & \nodata \\
6248 & Fe II & -0.7 & \nodata \\
6300 & [O I] & -1.2 & \nodata \\ 
6318 & Mg I & \nodata & \nodata \\
6363 & [O I] + Fe II & \nodata & blend$^{6}$  \\
6385 & Fe II & -0.8 & \nodata \\
6417 & Fe II & \nodata & \nodata \\
6433 & Fe II & \nodata & \nodata \\
6456 & Fe II & \nodata & \nodata \\
6516 & Fe II & \nodata & \nodata \\
6563 & H $\alpha$ & -267 & P-Cygni$^{7}$   \\

\tablecomments{Line identifications are tabulated for the strongest lines in NGC 346:KWBBe 200's optical spectrum.  Net equivalent width measurements are also 
cited for all lines which exhibit no significant evidence of blending.  
$^{1}$ blend of Fe II 4489+4491 \AA; $^{2}$ blend of 
Fe II 4520,4522 \AA; 
$^{3}$ blend of Ti II 4533 + Fe II 4534 \AA; 
$^{4}$ blend of Fe II 4549 + Ti II 4549 
\AA; $^{5}$ wings extend to -840 km s$^{-1}$, +1220 km s$^{-1}$; 
$^{6}$ blend of [O I] 6363 + Fe II 6369 \AA; 
$^{7}$ EW excludes Fe II 6516\AA. }

\end{tabular}
\end{table}

\clearpage

\begin{table}
\caption{Photometry and Summary of Fundamental Parameters \label{param}}
\scriptsize
\begin{tabular}{lc}
Parameter & Value \\
\tableline
\\ 
U & 14.834 $\pm$0.048 \\
B & 15.957 $\pm$0.048 \\
V & 15.606 $\pm$0.038 \\
I & 15.240 $\pm$0.148 \\
J & 14.551 $\pm$0.046 \\
H & 13.855 $\pm$0.043 \\
K' & 12.772 $\pm$0.032 \\
3.6 $\mu$m & 10.53 $\pm$0.05$^{1}$ \\
4.5 $\mu$m & 9.76 $\pm$0.04$^{1}$ \\
5.8 $\mu$m & 8.85 $\pm$0.04$^{1}$ \\
8.0 $\mu$m & 7.81 $\pm$0.03$^{1}$ \\
T$_{eff}$ &$\sim$19000 K \\
R$_{star}$ & $\sim$14 R$_{solar}$ \\
log (L$_{star}$/L$_{solar}$) & $\sim$4.4 \\
T$_{dust}$ & $\sim$800 K \\

\tablecomments{The optical and near-IR photometry for NGC 346:KWBBe 200 
has been extracted from the catalog of \citet{zar02}.  $^{1}$ The quoted 
Spitzer IRAC-band photometric errors do not include absolute calibration uncertainties, 
which are estimated to be $<$5\% \citep{ira06}. } 

\end{tabular}
\end{table}

\end{document}